\documentclass{kapproc} 

\usepackage[dvips]{graphicx}

\upperandlowercase

 \let\footnote\savefootnote

\kluwerbib 

\def\lesssim{\mathrel{\hbox{\rlap{\hbox{\lower4pt\hbox{$\sim$}}}\hbox{$<$}}}}
\def\gtrsim{\mathrel{\hbox{\rlap{\hbox{\lower4pt\hbox{$\sim$}}}\hbox{$>$}}}}

\newcommand{\ion}[2]{#1$\;${\sc{#2}}}
\newcommand{\hii}{\mbox{\ion{H}{ii}}}

\newcommand{\hb}{\mbox{H$\beta$}}

\newcommand{\sci}[2]{\mbox{$#1\cdot 10^{#2}$}}
\newcommand{\Ms}{\mbox{M$_\odot$}}

\begin{document}

\articletitle
{Massive Young Clusters}

\chaptitlerunninghead{Massive Young Clusters} 

\author{Jes\'us Ma\'{\i}z-Apell\'aniz\altaffilmark{1,2}}

\affil{\altaffilmark{1} Space Telescope Science Institute, Baltimore, USA \\
       \altaffilmark{2} Space Telescope Division, ESA, ESTEC, Noordwijk, Netherlands}

\begin{abstract}

	In the last decade we have come to realize that the traditional
classification of stellar clusters into open and globular clusters cannot be
easily extended beyond the realm of the Milky Way, and that even for our Galaxy
it is not fully valid. The main failure of the traditional classification is the
existence of Massive Young Clusters (MYCs), which are massive like Globular
Clusters (GCs) but also young like open clusters. We describe here the mass and 
age distributions of clusters in general with an emphasis on MYCs. We also
discuss the issue of what constitutes a cluster and try to establish a general
classification scheme.

\end{abstract}

\section{Introduction}

	The traditional classification for Milky Way stellar clusters is that
they are either globular or open. Globular clusters are 
old ($\sim 10$ Ga), massive ($\sci{3}{4}-\sci{3}{6}$ \Ms), metal-poor, and 
spherically-symmetric members of the Galactic halo. Open clusters are 
young ($\lesssim 1$ Ga), low-mass ($< \sci{5}{3}$ \Ms), metal-rich, and 
non-spherically-symmetric members of the Galactic disk. The increase in
resolution and light-gathering power provided by HST and the new generation
of ground-based telescopes has taught us that such classification is not
valid for other galaxies and that even for the Milky Way it is not completely
correct. Some clusters can be both young and massive and some galaxies can have
large numbers of such Massive Young Clusters 
(or MYCs, \cite{Maiz01b,Whitetal99b,LarsRich99}), which are the object of this
review. This paper is divided into three sections, each one of them 
corresponding to the three words that make up the name of these objects:
Massive (what are the masses of stellar clusters? how does the mass influence
the dynamical evolution of the cluster?), Young (what is the history of stellar
cluster formation? what do we know about the youngest clusters?), and
Clusters (what is the internal structure of a stellar cluster? how do we
classify them?). 

\section{Massive}

\subsection{Mass distributions}

	Recent studies have shown that the luminosity distribution of young
stellar clusters is a power law ($dN/dL_{cl}\propto L_{cl}^{\alpha}$) 
with $\alpha\approx -2$ over a large range of cluster luminosities
(\cite{Whitetal99b,Lars02}). On the other
hand, it has been known for some time that GC systems have log-normal
luminosity distributions. The mean value of the distribution for our Galaxy is
$M_V = -7.36\pm 0.17$ and most other well-studied galaxies have values in the 
range $M_V = -6.9$ to $-7.6$ (\cite{Harr91}).

	The masses of Galactic globular clusters are well known
(at least within a factor of two) since the 
resolution into individual stars allows for detailed dynamical modeling using 
radial velocities and/or proper motions (\cite{Meyl02}). Recent work using 
HST promises to improve the uncertainties (\cite{AndeKing03}). The mass 
function of the Milky Way thus measured also follows a log-normal distribution 
(\cite{FallZhan01}) with its peak centered at $2\cdot 10^5$ M$_{\odot}$. But 
what about MYCs, where most of the light is produced by short-lived stars which 
constitute only a relatively small fraction of the total mass? Do they 
have masses similar to GCs, implying that they have ``normal'' stellar IMFs 
that extend to low masses, or are they deficient in low-mass stars? The masses
of MYCs are harder to measure due to inadequate spatial resolution (implying 
the use of integrated data) and to the fact that the stars which
produce the most useful lines for measuring velocity dispersions, red
supergiants (\cite{HoFili96,SmitGall01}), are not present in the earliest 
stages of MYCs. Also, some clusters have double cores (\cite{Maiz01b}), others
are heavily extinguished, and some measurements of the
velocity dispersion can be affected by the presence of binaries
(\cite{Boscetal01}). Results show that most MYCs have indeed normal stellar 
IMFs and are as massive as GCs 
(\cite{HoFili96,Larsetal01}); some clusters may have
somewhat anomalous stellar IMFs (\cite{SmitGall01}) but the 
observational problems listed above could be a factor in those cases. 
Therefore, if the stellar IMF
is constant for all or most of the cluster-mass spectrum, the mass function of
young clusters (the cluster IMF)
should have the same functional shape as the luminosity
function, i.e. a power law with $\alpha\approx -2$.

\subsection{Cluster evolution and survival}

     Several processes drive the dynamical evolution of a
cluster: mass loss due to stellar winds and SNe, a process that heats the
cluster; two-body interactions that lead to relaxation, energy redistribution 
as a function of stellar mass, stellar ejections, evaporation, and possible 
core-collapse; binary star formation, another cluster-heating process; and
tidal interactions in the form of static tidal fields, tidal shocks, and
dynamical friction (all leading towards a faster
destruction of the cluster) and tidal merging (\cite{Gerh00}). These processes
are known to be tearing apart GCs little by little (\cite{Rocketal02}).

	How will these processes affect the evolution of MYCs. Will they become
the GCs of the future? \cite{FallRees77} determined that 
in order for a cluster to survive for a Hubble time, a large initial mass
($\approx$ \sci{3}{4} \Ms) was required. For lower-mass clusters the combined
effect of two-body interactions and tidal forces was too strong for the
cluster to last that long. Since we have seen that MYCs are indeed as massive
as GCs, this condition for long-term survival appears to be satisfied. However,
why is it that the mass function for young clusters is so different to the one
of GCs? The answer comes from the numerical simulations of \cite{FallZhan01}:
since the above mentioned processes destroy low-mass clusters more efficiently
than high-mass ones and since even the latter lose stars little by little, an
initial power-law mass function is easily converted into a quasi-log-normal one 
(actually, almost any reasonable initial mass function is converted into a
quasi-log-normal one). \cite{BoutLame03} backed some of these conclusions with
observations of 4 galaxies where the data can be explained if low-mass clusters
are preferentially destructed. Furthermore, those authors
observe that survival time scales 
are a function of the environment, as expected: the central regions of spiral
galaxies, such as M51 and M33, destroy clusters faster than in the solar
neighborhood, located at a larger galactocentric distance. A dwarf
irregular galaxy such as the SMC has an even more benign environment, due to
the weakness of tidal effects there. Therefore, a $10^4$ M$_{\odot}$ is
expected to last less than 100 Ma in the inner regions of M51 but could last
close to a Hubble time in the SMC.

\section{Young}

\subsection{History and triggering}

	Star cluster (of any mass) formation is a continuous process,
as the Galactic open cluster population shows, 
peppered with occasional bursts (\cite{BoutLame03}).
The formation history of massive star clusters is
harder to study due to their relative scarcity for young ages and to the
difficulties associated with measuring ages for unresolved systems
(\cite{KissP02}). GC systems show a color bimodality 
that can be interpreted as the result of two bursts of star formation 
but also as the merger (without new cluster
formation) of two preexisting galaxies; the latter model has some 
problems that make it less likely (\cite{KundWhit02}). The observed
new cluster population in current mergers such as the Antennae indicate that
this process is quite effective in forming considerable numbers of MYCs
(\cite{Whitetal99b}).

	MYCs appear to require galactic-scale massive events (such as galaxy
formation or mergers) to form in large quantities but more modest numbers can 
be produced without resorting to such processes. Thus, we have dwarf starbursts
such as NGC 4214, a Magellanic irregular where a few MYCs have formed along the
galactic bar in the last $\sim 10$ Ma (\cite{MacKetal00}). Even in the Local 
Group, where no major mergers have taken place recently, we have two good 
examples of MYC with ages $< 10$ Ma, 30 Doradus in the LMC and NGC 604 in M33,
plus a few other cases (\cite{Maiz04a}). A number of processes, such as
gravitational instabilities (caused by e.g. bars), massive-cloud collisions,
and galactic tidal interactions, appear to be able to produce massive star
clusters (\cite{TerE97}). The relative importance of each process has not been
studied in detail yet. Another aspect that probably deserves attention is an
in-depth analysis of the cluster IMF: is it identical for low-scale cluster
formation processes and for massive ones (the differences in the number of MYCs
originating only in a lower number for small-scale events due to the stochastic
filling of the cluster IMF) or are there intrinsic differences?

\subsection{Generations and the surrounding medium}

     	A compact low-mass cluster is born rather quickly
(few $10^5$ years, \cite{Bonnetal03}) and the core of a MYC is likely to form on
similar time scales. Therefore, MYC cores should be well approximated by
single-age populations. However, as we will see in the next section, many MYCs
have complex structures outside their cores which can be made up of several 
stellar generations. In 30 Doradus, the best studied MYC, five populations of 
different ages can be identified (\cite{WalbBlad97,GrebChu00}), from a 20-25 Ma
subcluster to an ongoing new generation; the core itself, R136, is 2-3 Ma old.
Some of those populations could actually be unrelated to the cluster (i.e. they
could be in the vicinity but not physically associated with it) but there
is good evidence that others are indeed part of the same cluster. Indeed, other
MYCs also have a second generation $\sim 3$ Ma older then the main star 
formation episode (\cite{Parketal92,WalbPark92}).

	The analysis of well-resolved MYCs shows that this presence of multiple
generations is related to the interaction between the cluster and the
surrounding medium (\cite{Maiz04a} and references therein). Massive Young 
Clusters are born from Giant Molecular Clouds and the ultraviolet
radiation from the massive stars with some help from stellar winds carves
an initial cavity in the molecular gas of a few tens of pc in size in the first 
$\approx 3$ Ma by first dissociating and later ionizing the molecular gas. The 
Giant H\,{\sc ii} Region around it is formed as a highly-stratified, thin 
($1-2$ pc) region on the surface of the Giant Molecular Cloud directly exposed 
to the UV radiation that can extend for several tens of pc
and that, in many senses, is nothing but a scaled-up version of what we observe
in lower-mass nearby \hii\ regions (\cite{Scowetal98,Ferl01}). This process also
drives shock waves into the molecular cloud, compressing the gas and triggering
the birth of new stars. After $\approx 3$ Ma, the first SNe start exploding 
and the resulting shock waves enhance the formation of this second generation.
However, they also contribute to sweeping away the molecular gas, so the
formation of new stars eventually stops shortly thereafter.

\section{Clusters}

\subsection{SSCs, SOBAs, cores, and halos}

    	Many stars are not formed in isolation but rather they are born in 
groups. Sometimes, the group is compact enough to be bound, at least for a 
period much longer than a typical orbital time for a given star, and we have a 
(real) cluster. In other occasions, the group is too extended and, although the
stars are born with similar velocity vectors (which differentiates them from 
nearby non-group members), the tidal field of the galaxy 
easily disrupts the group within one galactic rotation. In that case,
the group is called an (OB) association. It is not uncommon to have clusters 
(bound groups) within more extended associations, with both originating
from the same progenitor molecular cloud (\cite{deZeetal99}). This description 
was originally derived from the study of the relatively young low-mass clusters
and associations in the solar neighborhood. In the past decade we have found out 
that it can be easily extended to the upper end of the
young cluster mass spectrum. Thus, MYCs can be divided into two types:
Super Star Clusters (SSCs) are organized around a compact (half-light radius,
$r_{1/2} = 1-3$ pc) core while Scaled OB Associations (SOBAs) lack such 
structure and are more extended objects, with $r_{1/2} > 10$ pc 
(\cite{Hunt95,Maiz01b}). SSCs are bound objects and
represent the high mass end of young stellar clusters while SOBAs are (at least
from a global point of view) unbound and are the massive relatives of regular
OB associations\footnote{Note that by calling a SOBA a MYC we are introducing a
slight terminological inconsistency with respect to the low-mass end of the 
spectrum, since the classical definition of a stellar cluster implies a bound 
object. Here we are applying a less restrictive definition of cluster to
include associations and SOBAs, i.e. to mean a group of stars born from the same
molecular cloud within a short ($\sim 10$ Ma) period of time.}.
Furthermore, the core of some SSCs is surrounded by extended halos which are 
themselves similar to SOBAs in terms of structure and number of stars, thus
representing the high-mass equivalent of those associations with clusters
inside. All three types of MYCs (SSCs with and without halos,
and SOBAs) are well represented in the sample available in nearby galaxies 
(\cite{Maiz01b}).

\begin{figure}
\centerline{\includegraphics*[width=\linewidth]{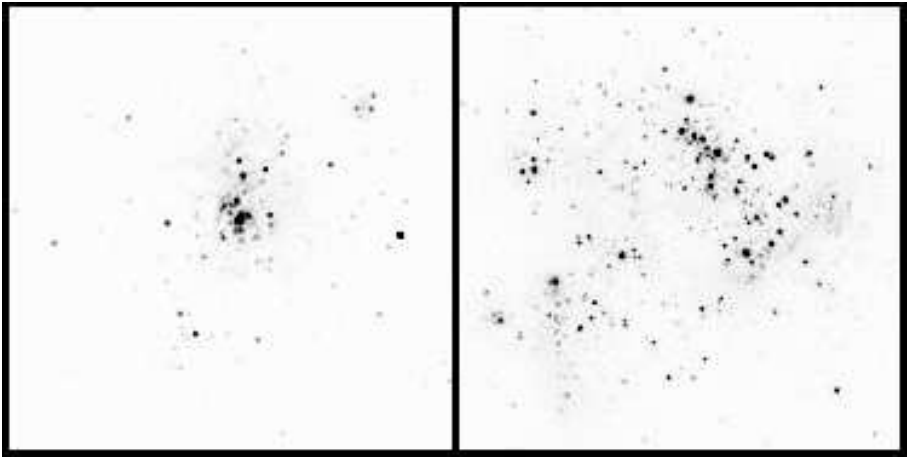}}
\caption{A comparison between 30 Doradus, an SSC with a halo (left), and 
NGC 604, a SOBA (right). Both
images were obtained using similar filters (continuum at \hb\ for 30 Doradus, 
WFPC2 F547M for NGC 604), and have the same physical size (120 pc $\times$ 120 
pc), orientation (N at top), and resolution (the ground-based 30 Doradus image
was degraded to attain this objective). R136, the SSC core, is the bright 
object at the center of the 30 Doradus image.}
\label{30dorngc604comp}
\end{figure}

	Why do MYCs come in these varieties? Part of the solution was discussed
in the previous section: the influence of an SSC in its surrounding medium can
produce, if sufficient material is available, a second generation of stars which
would form (part of) the halo. A more comprehensive explanation comes
from recent numerical simulations of galactic disks which show that 
molecular clouds are high-density, high-pressure regions that form mainly by 
turbulent ram pressure (as opposed to by self-gravity), which has its ultimate 
origin mostly in SN explosions (\cite{MacL04}). In this scenario, molecular
clouds are transient features which are easily created and destroyed.
Gravity would play a role only after turbulent pressure creates filamentary
structures dense enough to start collapsing (\cite{Bateetal03}). The 
simulations in those references are for ``normal'' conditions (leading to 
low-mass cluster formation) but the similarities that we have found between the
structural properties of low-mass and high-mass clusters and associations
suggest that MYCs may form in the same way, a hypothesis also supported by the
hierarchical nature of cluster formation (\cite{Bonnetal03}). The most important
difference between high-mass and low-mass clusters would be the need for a large
initial amount of gas and for an extremely high pressure in order to compress
it into a relatively small volume; such conditions could be caused by any of
the mechanisms (galactic collisions, gravitational instabilities\ldots)
discussed in the previous section.  
The hierarchical nature of the process suggests
that the same type of filamentary structures should form in the dense molecular 
gas during the early stages of formation of a MYC. Subclusters would then form 
along those structures and, if a region is dense enough to produce a large 
number of them within a small volume, a core would be formed when they merge.
The rest of the subclusters would form the halo or SOBA part of the MYC, with
the possible help of the shock waves created by nearby stars as described 
in the previous section. In
a time scale of the order of 10-30 Ma (the typical orbital periods around the
center of the cluster for stars located at radii of $10-20$ pc) the
relative positions of the stars there would bear little resemblance to their
original ones, but, for clusters younger than that, the halo could still have
some memory of the original filamentary structure of the molecular gas. Indeed,
some observations support this idea (\cite{Maiz01b,Maiz04a}).
Given the size of several tens of pc of the cloud, the whole
MYC formation process could take $\sim 10$ Ma from the time when the first 
stars are born until the time when the molecular material is dispersed, a value 
also consistent with the observed properties of well-studied MYCs.

\subsection{Size does matter: cluster survival and classification}

	As already described by \cite{FallRees77}, a high mass is not the only
condition necessary for long-term cluster survival: size does matter.
Clusters which are too compact are easily affected by two-body
interactions (though their immediate destiny is probably not destruction but 
only expansion) while clusters that are too extended get disrupted by tides, as
we have already mentioned. SSCs have the right intermediate size to
ensure survival (\cite{Maiz01b}) and one would expect them to become GCs in the
future. SOBAs, however, are too extended to survive for a long period of time
and are expected to dissolve rather easily and its members should become part
of the non-cluster population of their host galaxy. Intense star formation
episodes (i.e. starbursts\footnote{Note that this term is used sometimes to
denote only episodes of very high intensity and of galactic proportions.})
can produce either SSCs or SOBAs or both, so one should expect them to enrich
not only the massive cluster population of their host galaxies but also 
their field stellar population as a result of the dissolution of the SOBAs.

	What does all of this tell us with respect to the classification of
stellar groups? First, that the most clear division is not between open and
globular clusters but between real clusters and associations, both of which 
are the children of Giant Molecular Clouds, the first being bound and the
second unbound objects. Second, that real clusters only survive for a long time
if they are massive enough. Third, that SSCs are likely to become GCs after
some time. Indeed, it is probably a good idea to define an SSC as a cluster
which has the right size and enough mass to become a GC in the future. With 
those ideas in mind, I propose the following classification scheme:

\vspace{5mm}

\centerline{
\begin{tabular}{ccccc}
\cline{2-5}
 & \multicolumn{2}{c}{    Compact (bound)    } & 
 \multicolumn{2}{c}{$\;\;\;\;\;$ Extended (unbound)    } \\
 & Low mass & High mass & Low mass & High mass \\
\cline{2-5}
Young & Open & SSC      & OB association & SOBA \\
Old   &      & Globular &                &      \\
\cline{2-5}
\end{tabular}
}

\vspace{5mm}

Support for this work 
was provided by NASA through grant GO-09096.01-A from the Space Telescope 
Science Institute, Inc., under NASA contract NAS5-26555, and by the Spanish 
Government grant AYA-2001-3939.


\end{document}